\documentclass[sigconf]{acmart}
\usepackage{adjustbox}
\usepackage{graphicx}
\usepackage{wasysym}
\usepackage{tikz}
\usepackage{tabularx}
\usepackage{booktabs}
\usepackage{multirow}
\usepackage{array}
\usepackage{multicol}
\usepackage[xcdraw,table]{xcolor}
\usetikzlibrary{trees}
\usepackage{subfig}
\usepackage{url}
\newcolumntype{L}[1]{>{\hsize=#1\hsize\raggedright\arraybackslash}X}
\AtBeginDocument{%
  \providecommand\BibTeX{{%
    \normalfont B\kern-0.5em{\scshape i\kern-0.25em b}\kern-0.8em\TeX}}}

\setcopyright{acmcopyright}
\copyrightyear{2018}
\acmYear{2018}
\acmDOI{XXXXXXX.XXXXXXX}

\acmConference[Conference acronym 'XX]{Make sure to enter the correct
  conference title from your rights confirmation emai}{June 03--05,
  2018}{Woodstock, NY}
\acmPrice{15.00}
\acmISBN{978-1-4503-XXXX-X/18/06}

\begin{document}

\title{Investigating Novice Researchers' Perceptions of Research Privacy Within LLM-Assisted Workflows}
% Understanding Novice Researchers' Academic Privacy Concerns With LLMs

\author{Shuning Zhang}
\email{zsn23@mails.tsinghua.edu.cn}
\affiliation{
\institution{Tsinghua University}
\city{Beijing}
\country{China}
}

\author{Changxi Wen}
\email{wcx24@mails.tsinghua.edu.cn}
\affiliation{
\institution{Tsinghua University}
\city{Beijing}
\country{China}
}

\author{Eve He}
\email{eve.he@wisc.edu}
\affiliation{
\institution{University of Wisconsin-Madison}
\city{Madison}
\state{Wisconsin}
\country{U.S.}
}

\author{Ying Ma}
\email{ying.ma1@student.unimelb.edu.au}
\affiliation{
\institution{The University of Melbourne}
\city{Melbourne}
\country{Australia}
}

\author{Robert Xiao}
\email{brx@cs.ubc.ca}
\affiliation{
\institution{University of British Columbia}
\city{Vancouver}
\state{British Columbia}
\country{Canada}
}

\author{Xin Yi}
\email{yixin@tsinghua.edu.cn}
\affiliation{
\institution{Tsinghua University}
\city{Beijing}
\country{China}
}

\author{Hewu Li}
\email{lihewu@cernet.edu.cn}
\affiliation{
\institution{Tsinghua University}
\city{Beijing}
\country{China}
}

\renewcommand{\shortauthors}{Trovato and Tobin, et al.}


\begin{abstract}
Large Language Model (LLMs)-assisted scholarly workflows introduce critical privacy and intellectual property risks. As a uniquely vulnerable cohort driven by publication pressure and a lack of institutional support, novice researchers rely heavily on public LLMs, compelling them to navigate high-stakes privacy-publication trade-offs. To investigate these concerns, we conducted semi-structured interviews with 44 researchers across diverse disciplines. Our findings reveal that the fear of idea leakage paradoxically accelerates, rather than deters, reliance on LLMs, as researchers utilize them to expedite publication. They also held misconceptions that their ideas lacked the unique value to attract targeted attacks, and that their inputs would be safely diluted within massive datasets, preventing reconstruction. From interviews, we identified five types of mitigations including input fragmentation and adversarial probing, though we found that participants largely perceived these measures as ineffective. We outline implications including implementing institution-level sandboxed isolation, scenario-based privacy pedagogy, and verifiable data-deletion audits for transparency.
    % we outline three key implications for designing secure human-LLM interactions in research contexts, focusing on the adoption of privacy-preserving models, selective anonymization, and updated ethical guidelines for research.
    % The increasing use of Large Language Models (LLMs) in assisting novice researchers with ideation and conducting research, has introduced new challenges related to academic privacy. To understand what challenges exist and what are researchers' concerns, we first interviewed XX researchers, finding that XX and their main concerns. We then designed a large-scale survey (N=XX), letting them rate the acceptability of inputting these sensitive information. We found XXX. These results left three implications around privacy-preserving mode, selective anonymization and ethics, regarding academic research.  
\end{abstract}





\maketitle

\section{Introduction}

The rapid integration of Large Language Model (LLM)-based techniques into academic workflows has substantially improved the efficiency and scope of scholarly activities~\cite{grossmann2023ai,van2023ai}. However, this transformation has also introduced critical challenges. 
Specifically, the processing of unpublished ideas and proprietary datasets through third-party LLM systems can expose research materials not only to model providers but also to tool-mediated or third-party app ecosystems, where data collection and disclosure practices may be difficult for users to inspect~\cite{wu2025indepth}. This risk is widely discussed across the global scientific community~\cite{naddaf2025ai}.
% especially challenges regarding data leakage and the potential compromise of intellectual property, as is frequently echoed and widely discussed around the world~\cite{naddaf2025ai}. 

Previous research efforts have examined privacy concerns in human-LLM interactions for general end-users~\cite{zhang2024s,mireshghallah2024trust,zhang2025towards}, and built tools to help redact and sanitize content to preserve privacy~\cite{zhang2024adanonymizer,monteiro2025imago}. Researchers have also examined privacy violations and potential harms within educational contexts~\cite{kwapisz2024privacy,harvey2025don}. However, these efforts often center around general personal privacy and fail to address research privacy, a distinct domain involving specialized research ideas and contextual proprietary data. Furthermore, the human factors contributing to these security and privacy (S\&P) risks remain largely unexamined. We specifically focus on novice researchers, defined as individuals with limited publication experience~\cite{shah2009scientific}, typically graduate students~\cite{ellis2009towards}. They represent a critical demographic for three reasons: (1) as the primary authors responsible for drafting manuscripts and managing data, they handle the most sensitive intellectual property during the research process, (2) they are often active, ``early-adopters'' of AI tools and may more readily use AI tools to create new workflows, and (3) since they are still learning formal research norms, they may use these tools without fully recognizing the long-term privacy implications for their scholarly work. To address their challenges, we answer the following research questions (RQs):
% We specifically focus on novice researchers. As a demographic, they rely heavily on LLMs for research production but often lack formal training in academic data security, leaving their research workflows particularly vulnerable to emerging privacy threats. To address these challenges, we aim to answer the following research questions (RQs):

% while failing to address research privacy, which is a distinct domain involving proprietary data. Therefore, the specific practices of novice researchers, the extent of their exposure and the effectiveness of current mitigation strategies remain an open gap in the literature. Novice researchers are important as they have more needs on research production, have less training, and were in their relatively early stage of the research, who nees more value shaping. Towards these unresolved challenges, we aim to answer the following research questions (RQs):

$\bullet$ RQ1: What are novice researchers' mental models and privacy perceptions of LLM-assisted research workflows?
% research privacy concerns do novice researchers experience within LLM-assisted research workflows?
% What are the primary concerns of researchers regarding academic privacy when using academic chatbots?

$\bullet$ RQ2: What privacy protection practices do novice researchers adopt in LLM-assisted research workflows, and how do they perceive their effectiveness?

% $\bullet$ How can we enhance researchers' awareness of these concerns and empower them with tools to control their data?

$\bullet$ RQ3: What challenges hinder novice researchers from effectively protecting research privacy within LLM-assisted research workflows?
% do researchers face when attempting to protect academic privacy in LLM-assisted research workflows?

To answer these RQs, we conducted semi-structured interviews with 44 novice researchers across a range of disciplines. 

For RQ1, we identified that researchers valued the privacy of unpublished ideas and data. Specifically, they noted that ideas are more vague and can leak indirectly. Moreover, they hold misunderstandings around research privacy risks. They worried about the theft of unpublished ideas by service providers and other peers, causing professional consequences that hindered progress. Simultaneously, they underestimated the risks of model memorization and training, generally assuming that their data would be diluted within LLM training sets, or assuming that model refusals to output their sensitive data would protect their privacy.
% reconstructs and leaks unpublished methodologies. 

% For RQ2, we show that the privacy-utility trade-off is uniquely acute in researchers due to the contextual nature of intellectual property. Unlike standard \textit{Personally Identifiable Information (PII)}, masking core academic logic destroys the exact context LLMs require for meaningful output, frequently forcing researchers into a security surrender where core ideas are exposed to maintain tool utility. 
For RQ2, we found that novice researchers, operating without institutional sandboxes, employed ad-hoc mitigation strategies. They fragmented proprietary ideas or frameworks into different sessions or models to prevent data leakage, and used adversarial probing techniques such as asking whether LLMs memorize their data to empirically test the boundaries of profiling. However, they perceived these practices as mostly ineffective, and had to sacrifice effective assistance for privacy.
% To safeguard privacy, they are frequently forced to sacrifice the contexts that LLMs required to provide effective scholarly assistance.

For RQ3, we identified multi-faceted constraints. First, their anxiety about publication outweighs privacy needs, and the fear of idea leakage paradoxically leads them to use LLMs more. Second, infrastructure-wise, they lack private institutional sandboxes, having to rely on public LLMs with weaker privacy guarantees. Third, they lacked techniques for privacy management, such as data sanitation tools or formal AI privacy training. Finally, they lacked accountability, with little to no formal regulations on privacy, and limited ability to verify or audit the privacy protections offered by service providers, particularly for abstract ideas whose leakage is hard to track. These challenges call for actions such as institutional automated screening and separated tools, transparent risk visualization, and interactive privacy training.

Collectively, this paper's contributions are threefold:

$\bullet$ We characterized novice researchers' privacy perceptions, such as their emphasis on unpublished ideas and assumptions of privacy risks.

$\bullet$ We categorize five classes of privacy mitigation strategies such as data fragmentation and adversarial probing, and find that researchers largely perceive these mitigations as ineffective.

$\bullet$ We provide implications such as institutional automated screening and separated tools, transparent risk visualization, and interactive privacy training.

\section{Background and Related Work}

% We first discussed works around academic privacy, which is a closely related aspect of research privacy, then detailed works around LLM-powered academic chatbot. We finally reviewed papers around privacy-preserving LLM-powered chatbots.
% We first discuss works on academic privacy, which examines institutional and user-centric privacy challenges in educational technology. This literature establishes a parallel context that informs our focus on research privacy. We then review studies on LLM-powered academic chatbots, the technological context of our investigation, followed by existing privacy-preserving techniques for LLM-based systems.

\subsection{Research Lifecycle and Risks}\label{sec:rw_research_risk}

% Standard scholarly workflows comprise ideation, literature review, experimental design, execution, data analysis, and manuscript preparation stages. With the proliferation of LLMs, they face varying privacy risks when integrated with AI tools. These stage-specific vulnerabilities are summarized in Table~\ref{tbl:privacy_risk}. 
Standard scholarly workflows, encompassing ideation, literature review, experimental design, execution, data analysis, and manuscript preparation~\cite{creswell2019educational}, face diverse privacy risks when integrated with LLMs. These stage-specific vulnerabilities, summarized in Table~\ref{tbl:privacy_risk}, highlight a critical tension between these tools' utility and research privacy. Recent literature further quantifies these challenges, identifying issues such as confidential data exposure, intellectual property leakage, and insufficient transparency regarding data flows~\cite{shanmugarasa2025privacy}. Beyond technical leaks, researchers face ethical concerns such as bias, censorship, and fabrication, all of which require mindful engagement with generative tools~\cite{bjelobaba2025maintaining}.

\begin{table}[h]
\centering
\caption{Categorization of privacy risks across different research stages.}
\label{tbl:privacy_risk}
\small
\begin{tabular}{cc}
\hline
\textbf{Research stage} & \textbf{Privacy risks} \\ \hline

% 第一组：3行
\multirow{3}{*}{Ideation} & Leaked confidential ideas~\cite{bjelobaba2025maintaining} \\ %\cline{2-2} % 21
 & Systemic algorithmic bias~\cite{bjelobaba2025maintaining} \\ %\cline{2-2}
 & Restricted academic freedom~\cite{bjelobaba2025maintaining} \\ \hline

% 第二组：4行
\multirow{4}{*}{Literature review} & Copyright infringement~\cite{bjelobaba2025maintaining} \\ %\cline{2-2}
 & Confidential document exposure~\cite{heidt2024intellectual} \\ %\cline{2-2}
 & Research direction inference~\cite{lee2024paperweaver} \\ %\cline{2-2}
 & Extraction inaccuracy~\cite{bjelobaba2025maintaining} \\ \hline

% 第三组：3行
\multirow{3}{*}{Experimental design} & Research protocol exposure~\cite{bjelobaba2025maintaining} \\ %\cline{2-2}
 & Biased instrument design~\cite{bjelobaba2025maintaining} \\ %\cline{2-2}
 & Inaccurate informed consent~\cite{bjelobaba2025maintaining} \\ \hline

% 第四组：4行
\multirow{4}{*}{Data collection} & Participant privacy breach~\cite{bjelobaba2025maintaining} \\ %\cline{2-2}
 & Regulatory non-compliance~\cite{bjelobaba2025maintaining} \\ %\cline{2-2}
& Excessive data collection~\cite{sharma2024m} \\ %\cline{2-2}
 & Demographic quality disparity~\cite{bjelobaba2025maintaining} \\ \hline

% 第五组：4行
\multirow{4}{*}{Data analysis} & Raw data exposure~\cite{scuderi2026challenges} \\ %\cline{2-2}
 & Individual re-identification risk~\cite{ng2025analyzing} \\ %\cline{2-2}
 & Data integrity compromise~\cite{bjelobaba2025maintaining} \\ %\cline{2-2}
 & Model reverse engineering~\cite{shanmugarasa2025sok} \\ \hline

% 第六组：5行
\multirow{5}{*}{Manuscript preparation} & Unpublished result leakage~\cite{mann2025ai} \\ %\cline{2-2}
& Participant privacy violations~\cite{scuderi2026challenges} \\ %\cline{2-2}
& Intellectual property ambiguity~\cite{bjelobaba2025maintaining} \\ %\cline{2-2}
& Author identity leakage~\cite{bauersfeld2023cracking} \\ %\cline{2-2}
& Cybersecurity vulnerabilities~\cite{scuderi2026challenges} \\ \hline

\end{tabular}
\end{table}

These vulnerabilities are particularly acute in interactive chatbot deployments. Such environments amplify threats including the generation of fabricated content~\cite{li2024dawn}, operational vulnerabilities such as prompt injection~\cite{dong2025safeguarding}, and privacy risks involving intellectual property~\cite{chen2025survey}. Besides, while AI assistance dramatically boosts individual outputs by enabling researchers to publish 3.02 times more papers and receive 4.84 times more citations~\cite{hao2026artificial}, this reliance exacerbates S\&P vulnerabilities. Therefore, researchers widely discuss S\&P and safety risks of scientific tools despite acknowledging their transformative potentials~\cite{van2023ai,van2023chatgpt,wang2023scientific}. 
% Recent literature further quantifies the challenges of LLM-empowered science. Shanmugarasa et al.~\cite{shanmugarasa2025privacy} identify critical risks including confidential data exposure, intellectual property leakage, and insufficient transparency regarding data flows. Bjelobaba et al.~\cite{bjelobaba2025maintaining} delineate a broad spectrum of concerns such as bias, censorship, and fabrication, emphasizing the need for mindful engagement with generative tools. Their analysis across various platforms highlights potential ethical concerns relevant to different research phases.

% The integration of LLMs into academic workflows enhances productivity but introduces risks to privacy and research integrity. While researchers recognize the transformative potential of LLMs, they harbor persistent concerns regarding transparency and methodological rigor~\cite{van2023ai,van2023chatgpt,wang2023scientific}. In interactive chatbot deployments, these vulnerabilities are acutely amplified, manifesting as the generation of fabricated content~\cite{li2024dawn}, operational threats such as prompt injection~\cite{dong2025safeguarding}, and severe privacy risks that expose sensitive intellectual property~\cite{chen2025survey}. This tension between utility and risk is empirically evident. Although AI assistance dramatically boosts individual outputs by enabling researchers to publish 3.02 times more papers and 4.84 times more citations~\cite{hao2026artificial}, such heavy reliance exacerbates the core security and privacy vulnerabilities.

To better support this human-AI collaboration, methodological innovations rapidly evolved from single-phase, multi-agent ideation systems~\cite{baek2025researchagent} to adaptive writing assistants~\cite{nguyen2024human} and end-to-end autonomous research agents (e.g., \textit{InternAgent})~\cite{team2025internagent}. This progression reflects an escalating taxonomy of LLM autonomy~\cite{zhang2025evolving,zhang2025exploring}, prompting the academic community to establish strict ethical frameworks that mandate rigorous human vetting and transparent acknowledgment~\cite{porsdam2024guidelines}. 

\subsection{Research Privacy and Integrity}

While educational privacy primarily focuses on safeguarding student data and mitigating surveillance risks~\cite{kwapisz2024privacy}, research privacy involves protecting unpublished ideas, proprietary methodologies, and data integrity. Despite these distinct objectives, vulnerabilities observed in educational contexts can serve as a base for understanding the threats faced by researchers. Institutional hurdles, such as unprepared infrastructure and lack of standardized protocols, remain the primary barriers to data protection~\cite{gkrimpizi2023classification}. Resource constraints and limited internal expertise frequently force institutions to bypass rigorous security assessments~\cite{chanenson2023uncovering}, relying instead on trust-based vendor self-certifications~\cite{kelso2024trust}. This lack of oversight exacerbates accountability challenges, as educational institutions remain liable for vendor data mishandling, but have limited resources for contractual monitoring~\cite{rasner2021cybersecurity}. Even with established \textit{Data Protection Agreements}, inconsistent enforcement and insecure configurations persist~\cite{cohney2021virtual}, ultimately driving educators to adopt unsanctioned tools to meet professional demands despite acknowledging the associated privacy risks~\cite{kelso2025investigating}. 

Misaligned risk perceptions further complicate the educational privacy landscape. Parents often delegate privacy responsibilities to schools as they trust these institutions, creating a disconnect between perceived and actual data protection~\cite{zhong2023m}. Conversely, students report heightened privacy concerns regarding surveillance, like online proctoring, which undermine trust and provoke defensive behaviors~\cite{balash2021examining}. Furthermore, research indicates that students frequently overlook implicit data disclosure risks of assistive technologies, even when they remain highly cognizant of physical privacy threats~\cite{marsh2025don}.

Finally, evaluating educational technologies increasingly relies on nuanced risk assessments. As traditional privacy policies often obscure practical harms, studies used longitudinal user review analyses~\cite{yang2023discovering} and structured frameworks like the \textit{Analytic Hierarchy Process} to prioritize vulnerabilities such as algorithmic opacity and technical reliability~\cite{roy2024prioritizing}. While these methods evaluate general educational and institutional risks from GenAI systems, they overlook facets unique to researchers, such as threats to unpublished ideas and data.

\subsection{Privacy Perceptions and Thoughts}

We summarize existing work on S\&P perceptions towards LLM-assisted tools from end users and developers. We compare those with novice researchers, which have a hybrid role: they navigate the privacy landscape as end-users while managing data such as interview scripts within institutions.

For end users, privacy concerns span the entire data lifecycle. They navigate a privacy-utility paradox with LLM chatbots~\cite{zhang2024s,zhang2024ghost,zhang2025understanding}, actively seeking regulatory compliance and granular data control to establish trust dynamics across diverse platform ecosystems~\cite{ali2025understanding,liu2026designing}. Furthermore, users increasingly fear unauthorized extraction of their proprietary knowledge and configurations from AI agent creators~\cite{ma2025privacy,zhang2026characterizing,zhang2026privweb}.

Conversely, despite operating under strict regulatory frameworks, software developers frequently lack formal Privacy by Design (PbD) education and actionable guidelines~\cite{lee2024don,senarath2018developers}. Relying heavily on informal learning rather than standardized methodologies like anonymization~\cite{prybylo2024evaluating}, developers often treat privacy as a reactive, policy-driven compliance task rather than proactive engineering~\cite{li2021developers,lee2026privy}. Although usable security principles~\cite{acar2016you}, technology acceptance models~\cite{senarath2019will}, and internal ``privacy champions''~\cite{tahaei2021privacy} can promote structural privacy adoption, S\&P awareness and practices are still insufficient. Crucially, existing literature largely lacks research-specific privacy contexts. Established threat models for users and developers fail to capture the unique vulnerabilities of novice researchers, who protect unpublished conceptual frameworks, interpretive logic, and sensitive human-subject data against LLM memorization and leakage.

\subsection{Privacy-preserving LLM-powered Chatbots}

Recent studies identify a critical privacy-utility tension in LLM-powered chatbots. While users recognize data sensitivity, they often neglect proactive safeguards, exposing themselves to vulnerabilities such as insecure code generation~\cite{oh2024poisoned,tran2025understanding}. To mitigate these risks, contemporary research and product deployments develop various countermeasures. These solutions broadly follow two lines: architectural frameworks, such as localized and cloud-edge computing, and algorithmic safeguards, including cryptographic protocols and collaborative learning.

Architectural strategies reduce the attack surface by retaining sensitive data within local trust boundaries. Localized computing and on-device LLM inference minimize remote server transmission~\cite{xu2024device,gunter2024apple,jadhav2025edge}, while specialized tools reinforce this defense by enforcing transparent, on-device data minimization~\cite{zhou2025rescriber}. For computationally intensive tasks, cloud-edge collaborative systems dynamically route non-sensitive queries to the cloud while processing sensitive data locally~\cite{zhan2026prism}. Furthermore, hardware-backed Trusted Execution Environments (TEEs), such as Apple's Private Cloud Compute, extends these strict on-device privacy guarantees to cloud-assisted workflows~\cite{apple2024pcc}.

Researchers also proposed algorithmic protections. Cryptographic approaches, such as homomorphic encryption, secure data during processing~\cite{zhang2024privacyasst}. In collaborative settings, federated learning enables distributed model training across edge devices~\cite{zhao2024privacy}, supported by secure aggregation protocols that merge updates without exposing raw data~\cite{yan2024protecting}. Differential Privacy (DP) remains central to these frameworks, injecting calibrated noise into model updates and commercial workflows to protect individual records while maintaining model utility~\cite{liu2025urania,liu2025differentially,xu2025dp}.
% Central to these frameworks is Differential Privacy (DP), which injects calibrated noise into model updates and commercial workflows to safeguard individual records without compromising model performance~\cite{liu2025urania,liu2025differentially,xu2025dp}. 
Despite these advancements, deploying GenAI in research contexts introduces unique socio-technical complexities~\cite{prather2025beyond}. Literature lacks an understanding of how novice researchers navigate privacy risks, warranting investigation into their privacy needs and practices.
% A critical gap remains regarding how novice researchers navigate these privacy-utility tensions, necessitating the investigation into their ad-hoc mitigation behaviors and privacy needs.

\section{Methodology}

\subsection{Participants}

We recruited 44 participants (17 males, 27 females) with a mean age of 26.9 (SD=3.5, ranging from 22 to 38) through a mixture of personal contact networks (1 participant) and posting recruitment messages through LinkedIn (2 participants), RedBook (23 participants) and WeChat Moment (6 participants), and dedicated WeChat groups of different campuses (11 participants). Our participants included 3 assistant professors, 1 corporate researcher, 28 PhD candidates and 12 Master's students. Although the non-student participants had begun full-time research roles, they were still in the early stage of developing independent research trajectories, placing them within the novice researcher stage~\cite{wilson2021novice,aguCareerStages} (commonly referred to as ``recent graduate/postdoc'' or ``early-career'' stages). Participants represented a wide range of academic disciplines: 11 were from computer science, engineering, or related technical fields (e.g., Mechanical Engineering), 28 were from social sciences and humanities (e.g., Linguistics, Education), and 5 were from natural or life sciences. All participants reported having prior experience with LLM-based chatbots (e.g., ChatGPT, Claude, Gemini) for research-related tasks such as ideation and literature review. The interviews were approved by our university's Institutional Review Board (IRB), and participants were compensated 100 CNY each according to the local wage standard.
Participants' demographics are shown in Appendix Table~\ref{tbl:participant}.
% These participants included XX PhDs, XX post-docs, XX assistant professors, XX masters and XX bachelors. They are from diverse fields, including XX computer science researchers, XX social science researchers, and others from various disciplines. All participants reported having prior experience with LLM-based chatbots (e.g., ChatGPT, Claude, Gemini) for research-related tasks like ideation and writing literature review. The interviews were approved by our university's Institutional Review Board (IRB), and participants were compensated \$20 each according to the local wage standard.

% Our participants are balanced in their education levels, including a mix of master's/PhD students, postdocs and faculty/principal investigators. They are from fields with distinct privacy requirements, such as Computer Science/AI, Biomedicine and social sciences. We ensured all participants have prior experience using LLM-based chatbots (e.g., ChatGPT, Claude, Gemini) for academic workflows like ideation, literature review or code generation.

% We directly contact researchers through university departmental mailing lists, and professional academic networks. We distributed the posters online. We used a brief pre-interview screening questionnaire to filter candidates based on their research field. We filtered out those who was not researchers.

% Participants was compensated \$20 each according to the local wage standard. The study was approved by the our university's Institutional Review Board (IRB). 

\subsection{Interview Design and Procedure}

We designed semi-structured interviews to explore researchers' privacy concerns and mitigation strategies when using LLM-assisted research tools. The interview guide consisted of five sequential parts. (1) Background and tool usage: after introducing the interview, ensuring confidentiality, and obtaining consent, we asked participants about their experiences with LLM-assisted research tools (e.g., generative AI, cloud platforms, and code repositories). We gathered details on their usage history, primary use cases and examples. (2) Privacy perceptions and concerns: we explored participants' overall concerns when sharing research data with those tools. We asked them to define ``research privacy'', identify what information they consider sensitive, and explain how they developed these privacy concepts. They also discussed their experiences sharing research content with AI tools, and their comparative views on platforms provided by different countries. (3) Specific experience exploration: We guided participants to share experiences related to account linkage, unwanted identification, concerns about being observed, fears of research content leakage, from prior work in Sec~\ref{sec:rw_research_risk}. (4) Coping strategies: We asked participants to describe behaviors and strategies they adopt to mitigate these privacy risks, explicitly noting how they balance privacy concerns with research needs. (5) Finally, we asked participants for suggestions on how these tools could improve privacy protections. Interview scripts are shown in Appendix~\ref{tab:interview}.

We conducted interviews remotely via Tencent Meetings and Zoom. Session lasted between 31 and 58 minutes, averaging 40 minutes. Before starting, we explained the interview's purpose, ensured confidentiality, and obtained informed consent for both participation and audio recording. The interviews were recorded and automatically transcribed before analysis.
% During the interviews, one researcher led the conversation using the semi-structured guide, while another researcher took observational notes. % Upon completion, participants were compensated with \$20 Amazon gift card for their time.

\subsection{Data Analysis}

We analyzed the interview data using thematic analysis~\cite{clarke2017thematic}. All audio recordings were first automatically transcribed and then verified for accuracy by one primary author. Following that, four authors independently reviewed an initial subset of four randomly selected transcripts to generate initial codes. They then met to compare their codes, discuss discrepancies, and establish a unified codebook. After reaching consensus, the four authors applied the codebook to the remaining transcripts, iteratively updating the codebook, with each author coding 10 each. During this process, they intermittently discussed to resolve discrepancies. Finally, the authors collaboratively synthesized the codes into overarching themes and sub-themes. The final codebook is shown in Appendix~\ref{app:codebook}. Given the inductive and exploratory nature of our analysis, we chose not to calculate or report inter-rater reliability (as suggested by prior guidelines~\cite{mcdonald2019reliability}). We calculated code frequency post-coding. Below, participants are denoted using P1–P44. The original Chinese quotes were manually translated by one primary author fluent in both English and Chinese, and then checked by the other three primary authors, who were fluent in both English and Chinese for correctness.

\section{RQ1: Privacy Perceptions and Mental Models in LLM-Assisted Research}

Overall, we found participants primarily integrate AI into their workflows to support text and image-based \textit{manuscript preparation} (5/44 participants, e.g., P28), \textit{conceptual ideation} (21/44 participants, e.g., P30-31), and \textit{technical execution}, such as data analysis and coding (25/44 participants, e.g., P39-42). Additionally, they leveraged AI for \textit{information retrieval}, including literature synthesis (19/44 participants, e.g., P12-13) and addressing specific queries (6/44 participants, e.g., P23-24), as well as for \textit{evaluative tasks} like generating peer-review comments (3/44 participants, e.g., P26).

\subsection{Understanding of LLM Data Handling}

\textbf{Opacity and complexity.} 9/44 participants viewed LLMs as opaque ``black boxes'', citing a lack of transparency in internal data processing. This complexity makes the system's logic inaccessible to users. P28 expressed concern over this opacity, noting that \textit{``I do not understand the internal mechanics of this black box, yet its learning capacity appears remarkably potent.''} Similarly, P8 argued that, \textit{``it is impossible to know how submitted data is processed within the database or what conclusions are derived from it.''}

The perceived invisibility of AI operations diminishes user agency and fosters skepticism regarding privacy controls. Participants often viewed opt-out mechanisms or safety regulations as superficial, because the underlying data handling is intangible. P20 highlighted the resulting anxiety, noting, \textit{``when you lack sufficient confidence in the unknown, you experience anxiety and fear.''} P28 added that the elusiveness of the technology makes it hard to verify if restrictive settings are effective. 
% ... I cannot penalize the consequences if they occur.

% 9 and 5 should be selected in the tag"opaque and black boxes" one by one
5/44 participants also pointed out that these tools lack adequate privacy policies, hindering their understanding. Information asymmetry between corporations and users fostered distrust regarding data persistence and profiling. P7 observed that this complexity often leads to ``blind consent'', \textit{``everyone basically just scrolls to the bottom and clicks `Agree and Confirm' ... the algorithm is somewhat like a `black box'.''} 
% P4 linked this ``unknowability'' to a loss of sovereignty, \textit{``These rules are not something you can [control] ... my sense of vagueness and distrust stems from this.''}

Finally, 4/44 participants expressed concerns about the AI ``memory'' and the permanence of data in cloud infrastructure. The inability to verify data deletion lead to fears of commercial exploitation or profiling. P10 questioned the potential of entities to profit from the system's memory, noting \textit{``If someone were to buy its `memory' function ... could they profit from that?''} For non-technical users like P24, the continuous aggregation of data (e.g. via training) made it difficult to predict future risks.

\textbf{Unavoidable knowledge leakage.} 4/44 researchers believed that sharing data with AI models inevitably leads to leaks, rendering absolute confidentiality impossible. Distrusting corporate promises, many adopted a defensive mindset. As P6 remarked, \textit{``I haven't read their specific fine print ... I tend to assume the worst-case scenario regarding their intentions.''} 
% Therefore, users felt that standard security measures could not fully protect their research.

Unlike general data breaches, 8/44 researchers feared that the model will learn and reuse unpublished work. Participants viewed this as a form of intellectual theft. P25 argued that submitted information inevitably resurfaces, while P4 expressed concern that \textit{``ideas, thought processes, or the most basic research data could be grafted onto someone else's framework,''} 
% highlighting a critical lack of protection for academic intellectual property.

Finally, 7/44 researchers felt that these AI tools prioritize external interests over academic confidentiality. P24 observed that corporate policies are \textit{``likely influenced by factors like government regulation ... therefore, it isn't completely confidential.''} Therefore, users like P21 concluded that completely avoiding these tools was their only safe option.
%7 includes two tags

\textbf{Iterative optimization.} 8/44 participants perceived their data as a functional asset essential for model refinement and personalized experiences. This data-for-service exchange is considered a necessity to drive model optimization and enhance system memory regarding user preferences. For some, the benefits of rapid feedback and improved performance outweighed privacy costs. P6 noted, \textit{``Their using it for training doesn't conflict with my interests. My goal is to get quick feedback.''} Consequently, users recognized that the utility brought by LLM-assisted tools is linked to the learning process where the system enriches its understanding through inputs (P24).

% Perceived privacy risks were often mitigated by participants' belief that individual contributions are anonymized or rendered insignificant when aggregated into massive training corpora.
On the contrary, 3/44 participants held the belief that individual contributions are anonymized when aggregated into massive training corpora, mitigating their perceived privacy risks. Participants frequently suggested that the scale of data synthesis grants them a form of collective invisibility. P6 expressed comfort in this anonymity, stating, \textit{``I find it acceptable because I feel I am linked with a larger collective ... I become invisible within that group, so I am not worried.''} This view is further echoed by participants' assumption that LLMs primarily rely on public information rather than ongoing research or unpublished ideas for training (P36).

Despite the benefits, 21/44 participants expressed concerns regarding IP leakage or unintended knowledge transfer during the model optimization process. 14/44 participants expressed anxiety that unique insights provided by one user can be inadvertently retrieved and presented to others as synthesized output (P7). P4 questioned \textit{``whether our proprietary ideas might eventually become one of the data source it masters and be provided as feedback to others during interactions.''}
% This concern centers on the AI's role as a knowledge conduit, where data stored in global databases may be retrieved to assemble plausible answers using original user content (P7).

\textbf{Obfuscation by AIs.} Participants viewed privacy protection as a process of semantic transformation, where the AI serves as a buffer by abstracting and fragmenting data. Participants thought that such abstraction detaches specific ideas from their owners rather than reproducing sensitive information, especially finding that the system also provides fragmented viewpoints than reproductions of user projects (P23). P25 described this as a form of protection, noting that \textit{``AI receives and re-interprets information, which prevents the intentional leaks associated with human handlers''}. 
% P23 added that the system provides fragmented viewpoints than identical reproductions of user projects.

Furthermore, participants believed that automated data pipelines minimize privacy risks by removing human intervention (P9). The institutional division of labor suggests that no single group can aggregate or access comprehensive profiles. P9 argued that, \textit{``programmers lack the ability to scrape information across disparate departments.''} 
% Within this automated framework, they assumed that the system performs selective filtering, processing inputs as speech and retaining only data required for internal use (P31).

Finally, 2/44 participants perceived the model's refusal mechanisms as the primary system-level mitigation strategies. P12 noted that \textit{``these guardrails allow the model to piece together a puzzle based on the information provided, than actively pry into others' privacy.''} This structural vagueness was perceived as a protective layer that ensures anonymity while delivering detailed insights.

\subsection{Perceived Privacy Risks}

\textbf{Valuable information.} Participants enumerated various types of important information. First, participants fear that unique inquiries could lead to academic scooping (P1,P9-10,P13,P19-20,P27,P36). P28 observed, \textit{``sometimes the thinking or logic this AI speaks of is so consistent and similar to what my labmates say.''} Second, uploading unpublished manuscripts for proofreading creates a tension between editing utility and data retention risks (P7,P26,P35,P42). P23 admitted, \textit{``you can use this advice to revise ... where I would upload my entire, completely unpublished first draft.''} Third, raw empirical data like field notes and medical records are viewed as non-replicable competitive advantages (P4,P6,P31). P8 stated, \textit{``the valuable things are ... the exclusive empirical materials I have gathered.''} Fourth, risks include biometric leakage and the exposure of human subject identities due to improper de-identification (P3,P12,P20,P31,P35). P27 confessed, \textit{``I don't perform pre-processing on the transcripts or handle de-identification.''} Finally, researchers also view iterative prompting strategies and restricted institutional materials as private intellectual assets (P6,P10). 
% P6 remarked, \textit{``I don't want others to see my workflow.''}

\textbf{Unauthorized disclosure of intellectual property.} 21/44 participants worried about the unauthorized disclosure or external acquisition of novel concepts and unpublished manuscripts, which could lead to scooping and loss of academic standing. They feared that submitting core assets to AI models might expose their ideas to competitors, thereby causing idea convergence. Users also perceive a misalignment between research value and corporate profit motives, exacerbating the risk of intellectual property leakage by service providers.

\textbf{Empirical data exposure.} 4/44 participants highlighted the risks of uploading raw primary data, such as medical records and interview transcripts, which often lack confidentiality guarantees when processed by LLMs. Beyond raw data, participants worried that their unique interpretive frameworks could be exposed. Additionally, participants noted that the highly specific prompts used for literature reviews could inadvertently reveal sensitive, unpublished research directions and specialized knowledge. 

\textbf{Deanonymization and profiling.} 10/44 participants reported concerns regarding AI models' ability to reconstruct personal identities by synthesizing fragmented data points. This process acts as a narrowing mechanism where various clues are combined to uniquely identify an individual within a large population. P24 compared this to a statistical approach that \textit{``narrows the potential identity range until the user is exposed within that correct interval.''} Another participant recounted an instance where the model inferred their real name through semantic comparison, based on the provided conversation history (P27).

% The risk of spatial and professional deanonymization is amplified by the system's capacity to process multimodal inputs and niche search themes. 
2/44 participants thought that the system's capacity to process multimodal inputs and search themes amplified the risk of spatial and professional deanonymization. By analyzing visual clues in images or unique professional characteristics, AI tools can pinpoint residential addresses or locate specific scholars within narrow research fields. P24 described how the AI identified a specific apartment building and unit number from visual clues in a screenshot. Similarly, P31 noted that researchers in highly specialized domains are easily targeted by the system, due to the limited number of active practitioners. %P24 31

Furthermore, 4/44 participants thought that continuous interaction enables AI systems to implicitly accumulate data and construct detailed user personas encompassing sensitive behavioral and psychological traits. P20 expressed discomfort with this covert profiling, noting that the AI accurately inferred their age range despite never being explicitly provided with demographic details. Models are reported to also deduce their hobbies, research focus and personality traits (P3-4). This inferential capability creates a tension between the utility of personalized assistance and the necessity of session isolation, where context retention may introduce privacy degradation and cross-session bias risks (P3,P35). Capturing this discomfort with profiling, P4 reflected, \textit{``it told me a lot, what I like, what my hobbies are, what my focus is, what kind of personality I roughly have, so I think it is like an observer.''}
%P4 9 20 27

Finally, 2/44 participants thought cross-platform data synchronization within corporate ecosystems brings additional risks. Integrating LLM conversational logs with broad e-commerce or social media networks create a pervasive targeted advertising environment. P8 feared that conversational data could be repurposed for shopping predictions within a large corporate network. This cross-device profiling leaves users feeling personally located and granularly annotated (P6).

\textbf{Memorization and data deletion.} 12/44 participants expressed concerns that sensitive research inputs might be permanently integrated into the model training corpus or internal memory, leading to involuntary data reuse (P20,P36). They feared that the retained sensitive information might be invisibly abused (P27,P35), and AI systems could directly learn from uploaded images or other personal content without proper anonymization (P40,P42). Some noted that AI's strong learning capabilities amplify these threats (P3-4). P31 questioned, \textit{``there will always be a memory, right? So then you worry about whether this memory will become public.''} 
% (P31). 
% For instance, 

4/44 participants held distrust regarding data deletion mechanisms, frequently observing cross-session memory retention (P7,P10, P36). Many users discovered that clearing a conversation thread from the user interface did not erase corresponding memories from the backend (P7,P10). Users feared that the synthesis of professional habits and domain-specific knowledge could compromise intellectual property (P31). They viewed the longitudinal memory of AI systems with deep suspicion, fearing the commercial exploitation and unauthorized monetization of their profiles. P10 expressed this apprehension by stating, \textit{``if someone bought its memory function or something, could they also profit from it?''} They also recounted instances where the model proactively referenced past research ideas in entirely new dialogue windows, even months after the original session was supposedly deleted, and personalization settings were disabled (P36). Illustrating this illusion of user control, P10 noted, \textit{``I would delete it, but I feel that after deletion, it still remembers these things.''} (P10)

\textbf{Security vulnerabilities.} 4/44 participants thought there are multiple technical and operational vulnerabilities that facilitate unintended data exposure, including API key leakage, breaches in third-party integrations, and accidental privacy triggers. They feared about improper access management, particularly within shared account environments or on multiple devices (P6,P12). Users noted that failing to terminate sessions allows unauthorized individuals to invisibly retrieve historical queries and misuse sensitive information (P12). Highlighting this physical access risk, one participant warned, \textit{``if we forget to log out, and someone else uses my computer, they could also see these private things.''} (P12)%P6 10 12 31
% Beyond immediate risks, \textcolor{red}{XX/XX} participants 

\textbf{Research vs. daily privacy risks.} While participants detailed privacy risks of LLM-assisted research tools, 8/44 participants expressed no concerns for data leakage, and believed that the risks for research privacy were lower than for everyday interaction. They mentioned that AI-induced risks are minimal compared to existing digital threats or academic norm violations (P4,P42). They explained that most ideas are largely similar, and they do not think they possess unique ideas that others could not come up with. Furthermore, they felt that LLMs were generally not capable of providing help detailed enough to support their specific ideas, and instead restricted their inquiries to their broader research aims (P25,P42). Stepping back, they thought that LLMs are usually trained on massive datasets. They therefore believed that it is hard for LLMs to memorize specific user-inputted content (P4,P6,P39). Still, participants believed the reduction of privacy risks relied on contractual behavior (P20,P25).

18/44 participants drew a clear distinction between the institutional governance of research data and the perceived degradation of personal privacy in daily lives. They noted that research privacy is strictly regulated by ethical mandates and institutional review boards, whereas daily personal data is increasingly viewed as beyond individual control (P13,P29). Due to the frequency of data breaches, many participants adopted a fatalistic perspective regarding their everyday data, concluding that individual defensive measures are often futile against data collection (P29). As P13 noted, \textit{``In research context, privacy considerations are primarily framed within the scope of ethical standards and institutional review boards.''} (P13) 
%1 3 4 6 7 10 13 20 23 26 27 31 35 36 37 38 39 41 

Conversely, 8/44 participants argued that the repercussions of research privacy failures are more severe than those occurring in daily lives. They suggested that while daily privacy leaks are concerning, they lack the immediate impact on professional survival that research leaks entail. As P26 explained, \textit{``the consequences of research data leakage are immediate and severe. It involves either substantial legal penalties or losing the priority of discovery.''} Consequently, participants reported employing selective disclosure strategies because their professional lives are based on the confidentiality of their research findings (P20).
%1 4 7 10 26 27 37 39 

2/44 participants explicitly prioritized daily privacy, citing its direct implications for physical and financial risks. They noted while a research idea's value remains speculative prior to publication, exposure of personal identifiers, especially those of participants, yields concrete harms (P6,P23). P23 stated, \textit{``these data points are inextricably linked to physical safety, financial security, and property rights.''}%P6 23
% They emphasized that the exposure of personal background or financial credentials directly compromises the stability of their daily lives (P23).

Despite these contextual differences, 6/44 participants expressed a reluctance toward unauthorized information disclosure. They perceived that privacy is defined by an individual's choice to withhold information, regardless of whether that information relates to professional output or personal habits (P4,P31). P31 observed, \textit{``in a research context, one is naturally reluctant to disclose a nascent idea.''} (P31) P4 concluded that \textit{``unauthorized disclosure of private information is universally undesirable because privacy ultimately implies information that I explicitly chose not to share.''} (P4)
%P3 4 9 24 31 36

\subsection{Knowledge Sources}

\textbf{Formal pedagogical and institutional instruction.} 6/44 participants reported lacking organized mentorship. While some participants received guidance from advisors or online, most participants rarely acquired privacy knowledge through formal pedagogical sources, often finding institutional instructions inadequate or inaccessible. Furthermore, existing educational courses were frequently criticized as flawed. They noted that online assessments could be easily bypassed through manual guessing (P31). \textit{``Ultimately, participants can pass by memorizing the correct answers revealed after initial correct attempts.''} (P31) While a minority of participants encountered rigorous data privacy coursework within their institutions, such resources were typically restricted to computer science disciplines and inaccessible to the broad researcher community (P31).

\textbf{Empirical engagement and experiential discovery.} 9/44 participants developed their understanding through empirical discovery. This includes direct interaction with LLM-assisted systems, where they employ trial-and-error and self-directed inquiry to navigate technical complexities. For instance, P39 acquired direct insights into data confidentiality protocols through extensive experimental work. Similarly, P3 cultivated privacy awareness by manually exploring settings and meticulously reviewing service agreements.
%P3 4 20 23 26 31 36 37 39
Professional research and external context also play a role in their discovery. P26's perspective on AI governance was informed by professional research and engagement with related literature (P26), while P4 gained insights into the competitive nature of the field by observing research publication cycles and risks of preemptive publication. For others, industry experience and media analysis fostered their awareness (P20). As P20 noted, \textit{``my concerns were shaped by firsthand experience during an internship at a LLM company''} (P20).

\textbf{Peer-mediated and collaborative discourse.} 8/44 participants favored peer-mediated communications, where knowledge is disseminated through informal professional networks and interpersonal exchanges. P25 observed that direct scholarly discussions regarding privacy remain relatively sparse, with discourse instead centering on pragmatic security measures such as account protection. This focus is particularly relevant in collaborative settings where, as P25 noted, \textit{``multiple users may inadvertently gain visibility into sensitive logs.''} 
%P4 7 25 34 36 37 38 43 

\textbf{Scholars and media.} 11/44 participants gained knowledge around risks by synthesizing information from academic literature, public reporting, and social media platforms (P4,P7). P6 and P37 identified persistent concerns regarding intellectual property rights and the impersonation of users on social media to solicit funds (P6,P37). For others, awareness was reinforced by specific case studies involving data leakage during AI interactions (P7,P20). As P20 remarked, \textit{``A query regarding a specific assignment reportedly prompted AI to output an entire completed document, including sensitive personal identifiers.''} (P20). P4 further emphasized that this perspective is informed by a synthesis of diverse sources including specialized social media channels and scholarly articles.
%3 4 6 7 13 20 23 26 31 37 38

\section{RQ2: Research Privacy Protection Practices and Perceived Efficacy}

Participants took different ad-hoc practices to protect their privacy, as shown in Table~\ref{tab:complex_mapping}, which they often perceive as ineffective.

\begin{table*}[htbp]
\centering
\caption{Map of researchers' privacy risks (RQ1) to mitigation practices (RQ2) and challenges (RQ3), according to participants' own mentions. Filled circles indicate associations mentioned by participants, while open circles indicate no association.}
\label{tab:complex_mapping}
\small
\newcommand{\tiltcolumn}[1]{\adjustbox{angle=45,lap=\width-0.5em,margin=0 0 0 0}{#1}}
\begin{tabularx}{0.85\textwidth}{>{\raggedright\arraybackslash}p{0.32\textwidth} cccccc @{\hspace{10em}} cccc}
% \toprule
 & \multicolumn{6}{c}{\textbf{Practices \& Mitigation (RQ2)}} & \multicolumn{4}{c}{\textbf{Challenges (RQ3)}} \\
 \cmidrule(lr){2-7} \cmidrule(lr){8-11}
\textbf{Privacy Risks (RQ1)} & 
\tiltcolumn{Architectural isolation and access control} & 
\tiltcolumn{Data sanitation and obfuscation} & 
\tiltcolumn{Fragmentation and decoupled processing} & 
\tiltcolumn{Adversarial testing, probing and deletion} & 
\tiltcolumn{Human-centric verification and awareness} & 
\tiltcolumn{Ineffectiveness of current mitigation} & 
\tiltcolumn{Privacy-publication trade-off} & 
\tiltcolumn{Intransparency and informational asymmetry} & 
\tiltcolumn{Lack of control} & 
\tiltcolumn{Accountability deficit} \\
\midrule

Unauthorized disclosure of intellectual property & $\Circle$ & $\Circle$ & $\CIRCLE$ & $\CIRCLE$ & $\Circle$ & $\Circle$ & $\CIRCLE$ & $\CIRCLE$ & $\Circle$ & $\Circle$ \\

Algorithmic memorization & $\Circle$ & $\CIRCLE$ & $\Circle$ & $\Circle$ & $\Circle$ & $\CIRCLE$ & $\Circle$ & $\CIRCLE$ & $\Circle$ & $\CIRCLE$ \\

Empirical data exposure & $\Circle$ & $\CIRCLE$ & $\Circle$ & $\Circle$ & $\CIRCLE$ & $\Circle$ & $\Circle$ & $\Circle$ & $\CIRCLE$ & $\CIRCLE$ \\

Deanonymization and profiling & $\Circle$ & $\CIRCLE$ & $\CIRCLE$ & $\Circle$ & $\CIRCLE$ & $\Circle$ & $\Circle$ & $\Circle$ & $\CIRCLE$ & $\CIRCLE$ \\

Security vulnerabilities & $\CIRCLE$ & $\Circle$ & $\CIRCLE$ & $\Circle$ & $\Circle$ & $\Circle$ & $\CIRCLE$ & $\CIRCLE$ & $\Circle$ & $\Circle$ \\

Risk assessment skepticism and negligibility & $\Circle$ & $\Circle$ & $\Circle$ & $\Circle$ & $\CIRCLE$ & $\CIRCLE$ & $\CIRCLE$ & $\Circle$ & $\Circle$ & $\CIRCLE$ \\

\bottomrule
\end{tabularx}
\end{table*}

\subsection{Architectural Isolation and Access Control} 

9/44 participants prioritize environmental isolation to prevent data from leaving private boundaries, often shifting from cloud-based infrastructures to localized deployments. 6/44 participants employ technical controls, most notably by disabling model training and data-sharing functionalities. These efforts are complemented by revoking unnecessary system permissions (P7-8,P23,P27,P36-37) and using session-specific controls, such as incognito windows, to ensure immediate data deletion (P28,P36). For highly sensitive workflows, participants adopt hybrid models, using localized models for prototyping/analysis, and remote models for other tasks to mitigate external exposure (P29). In stances of extreme concerns, however, researchers may avoid using AIs at all to guarantee data integrity (P37).
%p6 7 8 23 24 26 27 29 36

3/44 participants manage privacy through identity isolation. They avoid credential sharing to maintain individual account integrity (P10,P25,P28). However, they also intermittently engage in informal sharing practices, which complicates governance and compromises security (P25). Furthermore, premium subscriptions and platform reputation serve as proxies for trust. They perceive paid accounts to offer protections for high-stakes manuscripts (P36,P42), which lead users to favor reputable first-party providers with established infrastructures (P25). They actively avoid platforms known to use user interactions for model training (P8).

Despite these measures, participants emphasized various barriers. In particular, novice researchers lack the resources for localized deployments. They also mentioned constraints like high capital costs (P36), inferior performance compared to frontier cloud models (P8), and fragmented user experience (P23). These behaviors are also different across disciplines, with researchers in STEM fields having more privacy-preservation attempts and practices than those in the humanities (P24). Finally, many researchers remain doubtful as to whether their data is truly excluded from backend databases, regardless of the localized settings or controls applied (P7,P36).

\subsection{Data Sanitation and Obfuscation}

15/44 participants sanitized their data prior to input. Some exercised data minimization, deliberately restricting the volume and sensitivity of their inputs to reduce potential exposure (P28,P31). Others used strategic prompting to extract analytical insights, concealing sensitive context to avoid full data uploads (P42).

Beyond volume reduction, participants employed specific obfuscation techniques to alter the data. Some have well-defined de-identification ``strategies'', such as redacting personal identifiers and geographical locations before transmission (P26). They noted that properly anonymized data makes it technically infeasible for the system to reconstruct respondent profiles (P13). To balance anonymization with the contextual integrity for accurate AI outputs (P7), users replaced specific data with generalized placeholders and rephrased their inquiries (P4,P8). They also routinely withhold key details from their prompts (P4).

Despite these measures, participants mentioned the limitations of manual obfuscation. They felt that maintaining data control requires substantial cognitive effort, which makes some give way to complacency (P26). Furthermore, users expressed concerns regarding model capabilities, fearing that AI's cross-session memories could aggregate fragmented inputs over time (P20).

\subsection{Fragmentation and Decoupled Processing}

7/44 participants employed fragmentation strategies to prevent models from reconstructing sensitive intellectual data. By decoupling complex tasks into isolated and non-contextual units, they thought that these methods obscured the scope of their research. Common approach include segmenting datasets and manuscripts prior to submission (7/44 participants), using multiple AIs for fragmented tasks (7/44 participants), and performing core work manually while reserving AI for final polishing (9/44 participants). P26 noted that \textit{``they only provide data fragments rather than the full dataset to avoid that the model access whole research idea.''} For highly sensitive inquiries, researchers also used locally deployed tools to process fragmented questions, aiming to mitigate exposure risks (P36).
%p26 10 9 7 24 34 36
% For instance, rather than uploading complete unpublished manuscripts, researchers processed isolated text segments to address specific stylistic elements, thereby denying the model access to the research (P24,P34).

% Instead of uploading complete unpublished manuscripts or raw datasets (P34), participants fragmented their documents. They submitted isolated text segments to refine specific grammatical or stylistic elements, thereby preventing the model from accessing the intellectual asset (P24). 
% P10 described a selective input strategy that extracts only specific segments required for summarization or planning, which \textit{``prevents the transfer of comprehensive datasets to the platform.''} P09 utilized a segmented approach, submitting small text passages to refine linguistic expression. P07 further explained that influenced by documented case studies, they now deconstruct manuscripts into isolated segments or individual sentences, which are then distributed across different AI platforms or multiple independent chat windows \textit{``to prevent any single model from reconstructing the coherent whole.''} P24 similarly partitioned texts into minor segments for targeted stylistic analysis without disclosing full documents, and P36 described using a locally deployed Copilot for fragmented inquiries when analyzing high-stakes research ideas.

Participants distributed these fragmented tasks across multiple chat sessions to reduce risks (P35). By dispersing their prompts, they ensured that no single system retained a comprehensive overview of their work (P7). Additionally, users put prompts on multiple LLM-assisted research tools to mitigate identity reconstruction and avoid verbatim replication (P35). They submitted identical prompts to various models and manually synthesized diverse outputs, creating composite texts with minimal similarity.

\subsection{Adversarial Testing, Probing and Deletion} 

4/44 participants conducted adversarial probing, role-playing, or testing. Some adopted an exploratory stance, trying adversarial probing against AI models to assess system boundaries. They instructed the AI to analyze their historical conversational data, to determine its automated profiling capabilities. P12 observed that the system initially exhibited built-in constraints against soliciting private data. With a similar method letting LLMs to summarize themselves, P36 conducted a comprehensive adversarial probe that yielded an alarmingly granular report. The generated profile accurately inferred their university affiliation, doctoral cohort, residential location, and recent psychological states entirely from prior dialogues. However, P36 noted that subsequent system updates appear to have strengthened privacy guardrails against such profiling.

Other participants employed deception or role-play to obscure risks. By constructing hypothetical scenarios based on real events, participants fed the AIs with synthetic data (P25). P25 thought \textit{``this ensures that the model cannot distinguish between factual data and hypothetical constructs.''} (P25) Additionally, participants also chose to manually rewrote the final outputs than adopting AI-generated text (P8).
% P8 described rewriting the final text manually rather than adopting AI's verbatim phrasing to \textit{``effectively bypass the risk of direct data exposure.''} P25 stated that scenarios presented to the AI were entirely synthetic and framed as simulations based on real events, a strategic use of role-play intended to obscure the underlying reality so that the LLM \textit{``cannot distinguish between factual data and hypothetical constructs.''}

Participants also requested data deletion, routinely deleting chat logs, specific keyword queries, and uploaded media immediately upon task completion, to minimize exposure windows and prevent risks associated with shared accounts (P6-7,P35,P39). This protocol was enforced most strictly when they inadvertently processed sensitive third party data (P20). While some users viewed UI-level deletion as a necessary security measure that provided psychological closure (P9), the majority remained skeptical of its technical efficacy. They empirically found that retroactive deletion cannot erase information already integrated into the global training corpus (P12), and strongly suspected that \textit{``deleted logs likely persist in the system background architecture.''} (P8) Furthermore, users expressed lingering uncertainty regarding the invisible retention of ambient background data captured during voice interactions (P20).

\subsection{Human-centric Verification and Awareness}

20/44 participants adopted human-centric practices to reduce privacy risks, ranging from avoiding AI use altogether and manually abstracting sensitive data to auditing model outputs, using step-by-step interaction protocols, and reverting to human-only workflows for sensitive tasks. 6/44 participants relied on themselves to think out ideas, avoiding AIs entirely. Instead of inputting raw or sensitive datasets into generative models, 12/44 participants manually abstracted the data prior to analysis (P31,P42). They meticulously audited AI outputs to ensure the models were not outputting personal arguments or unauthorized testimony, often through requiring explicit citations for cross verification (P12). Furthermore, 3/44 participants adopted granular, step-by-step interaction protocols. P4 described evaluating the model feedback at each incremental step, terminating the session if there are inaccuracies or privacy risks. They also retroactively reviewed their interaction histories to confirm that no personal identifiers or lifestyle data had been inadvertently disclosed to the system (P25).
%6：3 6 26 28 36 37
%12：31 26 13 8 7 4 25 24 36 29 42 34
%3： 4 25 12

Beyond output auditing, 3/44 participants frequently reverted to traditional human processes for highly sensitive tasks. To evade AI-generated text detectors, they manually restructured sentences to retain human writing patterns (P35). In collaborations, researchers noted that peers often explicitly stipulated that manuscripts undergo human-only editing without AI involvement (P19). For high-stakes discussions, participants preferred consulting senior mentors over using LLMs. They emphasized that established interpersonal trust and professional ethics provide a level of security against intellectual property theft that automated digital platforms cannot guarantee (P28).

4/44 participants highlighted the need for education and awareness. They thought that understanding underlying technical mechanisms, such as context windows and architecture limitations, was essential for formulating safe queries (P20,P28). However, P7 suggested that constant exposure to data harvesting in daily digital applications has eroded their privacy awareness. Consequently, participants stressed that \textit{``individuals must cultivate active vigilance, as one cannot rely solely on legislation for privacy protection.''} (P36)
% 1/44

\subsection{Ineffectiveness of Current Mitigation} 

8/44 participants expressed skepticism regarding the efficacy of existing privacy mitigation strategies. Participants viewed these protective actions as mere psychological comfort than robust security measures (P7-9,P12,P36). This skepticism is driven by their perceived inevitability of data harvesting and mistrust towards AI service providers. They held distrust towards the technological corporations operating these platforms, and expressed anxiety regarding the opacity of backend operations, and the potential for unauthorized surveillance. They argued that profit-driven entities lack the motivation to permanently erase valuable training data. As P4 explained, \textit{``my distrust is rooted in the unknowability and opacity of these systems.''} Participants felt entirely disempowered, lacking the technical resources to audit backend practices, and the legal standing to hold corporations accountable. As P29 noted, \textit{``I have a lack of trust in tech giants. They possess the capability to act as they please regardless of their public assurances.''}
% Despite these strategies, participants expressed uncertainty about how to protect their privacy. 
% Users acknowledged that their active mitigation efforts are confined to data entry point. Because the internal data processing mechanics of these algorithms remain inaccessible, they felt unable to verify whether their input-level protections successfully prevented downstream privacy violations. 

They also expressed concerns towaeds the discrepancy between user interface modifications and actual backend data persistence. They perceived user-side controls as superficial features that fail to prevent backend data retention (P10,P12,P31,P36), and therefore restricted their mitigations to the data entry point. As P25 noted, \textit{``I remain uncertain as to how these algorithms actually process my data or resolve the underlying challenges of privacy protection.''} Some suspected that deleting chat logs or disabling training permissions merely removes client-side visibility, while the underlying infrastructure retains the information (P10). As P7 noted, \textit{``the act of deleting or regenerating a dialogue may merely remove the information from the user interface. It remains highly probable that the data persists within the system.''} They argued that discarded inputs are actively integrated into the system's latent memory, similar to data recovery in hardware components (P12). 
% also cited analogies to hardware recovery to argue that discarded inputs remain fully accessible, and are actively integrated into the system's latent memory (PXX).

Furthermore, participants viewed data leakage as an unavoidable consequence of modern research workflows. As generative models have become indispensable, users felt compelled to accept the associated privacy risks with no viable alternatives (P37). Researchers adopted a dual-sided stance toward privacy loss, as P31 commented, \textit{``despite the risks, the utility of these tools necessitates their continued use.''}

% Given the perceived futility of these features, participants used mitigation strategy mainly to alleviate anxiety rather than achieve genuine security (P36).
To manage the lack of protection, individuals frequently rationalized their exposure by downplaying the value of their own intellectual property. They assumed that their routine research ideas or personal details lack sufficient value to incentivize corporate misappropriation. P12 stated, \textit{``I sometimes rationalize that as an ordinary individual without highly sensitive information, such stringent measures might not be strictly necessary.''}

\section{RQ3: Challenges to Protecting Privacy}

\subsection{Privacy-Publication Trade-off} 
8/44 participants highlighted the tension between the operational necessity of LLMs and the inherent anxiety regarding data exposure. Their primary concern was the potential for models to inadvertently leak unpublished research ideas, which paradoxically pressured users to accelerate their publication cycles to preempt AI-driven plagiarism (P20). Despite acknowledging the persistent threat of unauthorized data usage in model training (P36), participants overwhelmingly prioritized efficiency over strict data privacy, like using LLMs for labor-intensive tasks such as document translation, literature summarization, and image generation. Highlighting this dilemma, P20 explicitly noted, \textit{``between this productivity and anxiety, I still choose productivity.''} Consequently, users routinely accepted privacy compromises for utility, such as retaining sensitive chat logs for ongoing tasks (P35) or rationalizing data exposure as an unavoidable cost of operational convenience (P23). Ultimately, the competitive disadvantage of abstaining from AI tools was perceived as a far greater threat than the associated privacy risks (P36). As P23 stated, \textit{``despite knowing AI might leak my data ... the need to improve efficiency makes its use necessary.''}
%P20 4 6 8 19 23 36 35
% \textcolor{red}{XX/XX} participants prioritized functionality over privacy, viewing advanced AI models as essential for scholarly workflows. They accepted privacy risks as a trade-off for efficiency gains (P23,P35-36). Specifically, AI was used for labor-intensive tasks such as document translation, literature summarization, and image generation, as manual alternatives were considered substantially less effective (P19,P23,P35). As P23 stated, \textit{``despite knowing AI might leak my data ... the need to improve efficiency makes its use necessary.''}

This emphasis on productivity is driven by a competitive research environment that prioritizes execution speed over ideas' novelty. Although participants feared idea leakage, this concern paradoxically motivated them to use AI to accelerate their work, aiming to publish before their concepts could be exposed (P6,P8,P20). As P6 noted, \textit{``ideas are not precious now. The key is to quickly transform the idea into a paper through hard work. Speed is key ...''} Consequently, speed in the publication pipeline is deemed more critical than protecting concepts from AI scraping (P8). Reliance on AI efficiency has increased the risk tolerance, which is reflected in the fact that, users often bypass privacy protocols. For example, participants admitted to inputting confidential documents into unapproved models to prioritize speed over compliance (P6).

\subsection{Opacity and Informational Asymmetry} 

13/44 participants reported informational asymmetry regarding data collection and processing, leading to a demand for accessible privacy governance. Vague privacy agreements cause user anxiety concerning whether inputs are extracted verbatim or analyzed only as keywords (P23). Users criticized speculative terminology like ``using data'' and requested explicitly defined rights and responsibilities (P8). For non-technical users, ambiguous communication increases apprehension because trust requires demonstrated privacy (P36). As P23 summarized, \textit{``privacy policies need to be accessible to non-experts. We need to know exactly how our dialogues and documents will be stored or shared.''}
% Participants perceived an asymmetry in understanding how their data is collected, processed, and used for model training, leading to a strong demand for accessible privacy governance. Users frequently experienced cognitive dissonance when forced to accept vague privacy agreements, expressing anxiety over whether their inputs were extracted verbatim or merely analyzed for keywords (P23). They expected the boundaries of rights and responsibilities to be explicitly defined rather than speculative, criticizing broad terminology like ``using data'' as insufficient for informed consent (P08). Furthermore, users lacking technical backgrounds experienced heightened apprehension when communication remained ambiguous, emphasizing that establishing trust requires demonstrated data privacy (P36). As one participant summarized, \textit{``Privacy policies need to be accessible to non-experts. We need to know exactly how our dialogues and documents will be stored or shared.''} (P23)

Beyond textual policies, 13/44 participants advocated for operational transparency and engineering safeguards to demystify LLM logic. They proposed governance measures such as compartmentalizing database management from frontend algorithms to prevent unauthorized cross-departmental data extraction (P9). Despite lacking technical expertise, they expected protections against inadvertent third-party disclosure and continuous background learning (P24,P26). Platforms capable of proving intellectual property protection were viewed as viable for professional adoption (P24). Reflecting this need, P9 stated, \textit{``I advocate for operational transparency. For instance, I hope for a workflow where functional terms are strictly compartmentalized.''}
%P4 8 9 19 20 23 24 26 28 34 36 38 42 
% Participants desired clear internal data governance, suggesting the compartmentalization of database management from frontend algorithms to prevent cross-departmental data extraction (P09). Although acknowledging their lack of technical expertise regarding mechanisms like firewalls, users expected platforms to implement robust protections against inadvertent disclosure to third parties or continuous background learning (P24, P26). Platforms that can prove they safeguard intellectual property were perceived as significantly more viable for professional and academic adoption (P24). Highlighting this structural need, P09 noted, \textit{``I advocate for operational transparency. For instance, I hope for a workflow where functional teams are strictly compartmentalized.''}

The participants thought interface design is important for real-time risk disclosure and granular consent. They suggested comprehensive disclosure frameworks spanning onboarding, active usage, and post-usage phases (P4,P19). This includes safety assurances near input fields and specific protective options for sensitive topics or high-profile figures (P20,P42). They also valued clean frontend interfaces and proactive prompts for explicit consent regarding data learning (P8). As P28 argued, \textit{``The interface should offer more than just a simplified output. It should reveal the underlying reasoning process. Platforms should proactively solicit consent.''}
%TO BE SOLVED：the opinion of they valued clean frontend interface…… still keep unknow
% , asking at the outset

Finally, 4/44 participants highlighted the role of effective education. Effective training should include practical modules that cultivate awareness of proper engagement and data discretion (P31,P34). They also requested comprehensive manuals detailing the computational processing of sensitive information (P4). Highlighting the need for genuine educational value, P31 stated, \textit{``We need practical training modules rather than performative or superficial instructions.''}

\subsection{Lack of Control}
13/44 participants complained about lacking granular control in data management, often facing a binary choice between system utility and privacy. Users emphasized the necessity for session-specific controls, such as per-interaction toggles to withhold consent for model training (P7,P23). They desired selectable operational modes to maintain agency over sensitive research identifiers, without compromising tool functionality (P10,P41). To navigate these trade-offs, participants needed compartmentalized interfaces that separate persistent memory for ideation from ephemeral sessions for isolated tasks (P3). Highlighting this demand for granular autonomy, P23 stated, \textit{``for every new session, there should be a dedicated toggle to withhold consent for data disclosure or uploading.''} 
% P1 3 6 7 12 20 23 24 34 38 39 41 42 

Deep skepticism persists regarding current data retention policies, with users viewing frontend deletions as largely performative (P1,P20). Participants argued that once information is transmitted, its influence on the model is set, making superficial removals ineffective for protecting high-stakes intellectual property (P1,P42). Consequently, users advocated for verifiable and permanent deletion mechanisms, ensuring that purged data enters an irrecoverable state across all central databases (P10,P24). 
% Reflecting this pervasive distrust, P20 noted, \textit{``the system fails to execute a definitive purge in the absence of explicit user instructions.''}
% Users exhibited deep skepticism regarding current data retention policies, viewing local interaction log deletions as largely performative rather than offering genuine privacy benefits. Participants widely suspected that user data persists in backend databases despite removal from the frontend interface (P01, P20). They argued that once information is transmitted to the model, its influence is set, rendering superficial interface removal insufficient for protecting high-stakes research or sensitive intellectual property (P01, P42). Consequently, users advocated for the implementation of verifiable and permanent deletion mechanisms that ensure purged data enters an irrecoverable state across all central databases (P10, P24). Reflecting this pervasive distrust, P20 pointed out, \textit{``The system fails to execute a definitive purge in the absence of explicit user instructions.''}

Finally, participants advocated for data minimization and security protocol integration directly into AI models' architectures. Intrusive permission requests were viewed as irrelevant to core functionality and a violation of privacy prioritization (P38). Users insisted that systems should exclusively retain authorized information, demanding rigorous non-disclosure protocols and end-to-end encryption for confidential data sharing (P1,P38). Due to system opacity, researchers often opted for zero-retention modes featuring immediate session expiration to preempt unauthorized data resurfacing (P34,P39). Emphasizing developer responsibility, P42 argued, \textit{``it is imperative that platforms protect the data and intellectual property shared within the chat interface through enabling advanced security features.''}

\subsection{Accountability Deficit}
9/44 participants expressed a sense of powerlessness regarding their inputs, noting a lack of mechanisms to hold AI corporations accountable for data misuse. They viewed corporate S\&P disclosures as unreliable and opaque, which increased their anxiety about how information is processed. This lack of transparency creates an accountability gap, as they found it nearly impossible to seek legal recourse or organize collective action (P20). Despite recognizing these vulnerabilities, they remained highly dependent on these tools, leaving them in a compromised position. 

To mitigate this vulnerability, researchers advocated for state-level legislation and regulatory oversight, alongside practical remediation frameworks. For example, they prioritized institutional remediation, such as securing publishing rights following a leak, over monetary compensation (P19). However, others cautioned that aggressive regulation could stifle innovation (P28,P36). 
% For example, P9 emphasized the need for stricter enforcement: \textit{``I hope [regulators apply] criminal laws so someone truly pays the price ... [because] these companies may have already made a fortune.''} (P9)
% users also cautioned against extreme policy volatility, warning that overly aggressive regulation could stifle technological enthusiasm, leading some to rank personal caution as the most reliable immediate defense (P28, P36). Highlighting the demand for stricter enforcement, P09 noted, \textit{``I hope they could violate some criminal laws so someone truly pays the price, because as for money, I believe these companies might have already made a fortune.''}

Furthermore, users called for redefining legal privacy boundaries and mandating explicit consent for all data access. They argued that interaction logs should be treated with the same confidentiality as sensitive personal data, such as national IDs or financial records (P42). They criticized the unauthorized sharing of personal content and intellectual property, noting that automated systems often bypass consent and undermine user agency (P24,P35). P35 highlighted this loss of agency, \textit{``[The system] didn't respect me. It didn't ask. It didn't ask if I agreed to send my photos to others.''} As a result, they expected comprehensive regulations to govern data access and sharing. 

\section{Discussion}

\subsection{Research Privacy}

Our findings suggest that research privacy is different from daily, academic or corporate privacy. We contextualize these differences to highlight the unique vulnerabilities of the research lifecycle.

\textbf{Research privacy vs. daily privacy.} Research privacy diverges from daily personal privacy through its professional stakes and the complex nature of the underlying data. While daily privacy breaches often result in leakage harms or targeted advertising, the exposure of research data carries professional consequences, including the loss of research priority (i.e., scooping), and legal or ethical penalties. Unlike personal data, which users often perceive as having low utility for external adversaries, research data, comprising interpretive frameworks, unpublished manuscripts, and raw datasets, constitutes a high-value competitive advantage. This data is uniquely susceptible to involuntary memorization by LLMs and subsequent malicious reuse. 

Furthermore, research privacy is highly time-sensitive. Participants typically prioritize protection within the fixed window before publication. Unlike daily privacy, where removing explicit Personally Identifiable Information (PII) is often deemed sufficient for de-identification, standard masking is inadequate for research contexts. As research ideation relies on interconnected logic and unique methodologies, data remains attributable even after heavy obfuscation. These logical fingerprints allow LLMs to reconstruct intellectual contributions or perform granular profiling, rendering traditional data sanitation methods ineffective~\cite{zhou2025rescriber}. Therefore, while many users adopt a fatalistic view toward daily privacy due to pervasive tracking~\cite{xie2019revealing}, researchers maintain a high-anxiety stance toward their research privacy.

% While research and academic privacy share the foundation, their threat models differ. 
\textbf{Research privacy vs. academic privacy.} Academic privacy typically focuses on institutional surveillance and unconsented collection of educational data, such as coursework and student records~\cite{kwapisz2024privacy}. In this domain, the privacy-utility trade-off usually centers on balancing personalized learning benefits against data protection~\cite{khan2024teaching}. In contrast, research privacy centers on the misappropriation of IP by external actors and the leakage of novel concepts.

Despite the difference, both domains suffer from a deficit in privacy education. Beyond rudimentary compliance, such as signing consent forms for review boards, students and researchers both receive minimal training on modern cryptographic or technical data protection methods~\cite{kwapisz2024privacy,khan2024teaching}. This vulnerability is exacerbated by a lack of GenAI literacy even among faculty, particularly in non-CS disciplines, where advisors lack the technical expertise required to guide novice researchers through GenAI risks~\cite{bulut2026mapping,hassan2025awareness}.

\textbf{Research privacy vs. corporate privacy.} Research privacy is similar to corporate environments in the necessity of shielding sensitive data from competitors~\cite{ahmad2014protecting}. In industry, privacy is often treated as either a competitive advantage or a bottleneck to product development~\cite{tahaei2021privacy,li2021developers}. However, disparities exist in the underlying management infrastructure. Corporate entities may leverage enterprise contracts with LLM providers and the resources to deploy localized models to prevent leakage~\cite{udandarao2025democratizing}. Conversely, researchers lack the bargaining power to negotiate privacy agreements and the computational resource necessary for deployment.

The traceability of breaches also varies. While corporate leaks involving customer databases are generally well-defined and detectable via audit trails, leakages of research ideas are hard to trace and verify. Furthermore, the power dynamics of these environments lead to different privacy trade-offs. Corporate security protocols often achieve protection through invasive employee surveillance~\cite{stegman2023my,wilmot2026enterprise}. In contrast, the researchers we interviewed typically bear the privacy management burden themselves.

\subsection{Perceived Risks, Mitigation and Realities}

\textbf{Perceived risks vs. realities.} Researchers often underestimate their research privacy risks due to a dilution fallacy. Believing their individual ideas are safely hidden within massive datasets, they largely rely on manually removing identifiers. However, technical evidence contradicts this perception. Increasing a model's parameter size increases data memorization capacity~\cite{li2024llm} even despite corresponding increases in dataset size, making LLMs highly vulnerable to targeted data extraction~\cite{carlini2021extracting}. Furthermore, for high-stakes information, attackers may also force models to leak memorized data (e.g., IP-protected documents) using specific inducing techniques~\cite{duarte2024cop,cooper2025extracting,karamolegkou2023copyright}. Even manual obfuscation is structurally flawed: after partial redaction, structured data patterns still persist~\cite{zhang2025through,zhang2025pervasive}, allowing for reconstruction and even deanonymization~\cite{tari2026measuring}.

\textbf{Mitigation vs. realities.} To protect their data, users often adopt surface-level strategies like UI-level deletion, input fragmentation, and adversarial probing. Yet, these practices may fail to mitigate the risk. Deleting chat logs may provide a false sense of privacy. Because user inputs are algorithmically integrated into the model via training and memory mechanisms, frontend deletions cannot reliably reverse backend retention~\cite{li2026privacy}. Research on under-learning confirms that supposedly deleted data may leave exploitable residual traces~\cite{wang2025towards,naderloui2025rectifying,chu2025scrub}. Such sensitive data may expose models to membership inference attacks~\cite{zhang2025soft}. Similarly, input fragmentation is ineffective because network-level metadata can still expose hidden prompts and semantic intent~\cite{jeong2025network}. Finally, when a model refuses to disclose information, users often mistake this for data absence or effective privacy control. In reality, refusals typically reflect safety filters~\cite{gong2025safety}, and attackers can still use prompt injections to extract sensitive research context~\cite{hui2024pleak}.

\textbf{Regulatory misalignment.} Institutional regulations, such as IRB protocols, are not all compatible with AI-assisted research practices. As some universities rely on static compliance checklists, users find the guidance superficial and disconnected from their actual workflows. Additionally, institutions' research-related regulations are frequently fragmented: IRB protocols manage risks related to studies, IT departments oversee tool security and data leakage, PIs dictate workflow norms, and research offices govern IP ownership. In practice, novice researchers need to manually synthesize these rules to determine if a specific AI interaction is compliant. Without a unified guidance framework, users form ad-hoc mental models to navigate the system's opacity~\cite{ma2025privacy}. This gap between technical reality and lay mental models renders existing policy communication highly ineffective~\cite{van2024department} and undermines privacy-preserving behavior~\cite{yang2025privacy}. This failure stems from the mismatch where governance bodies assume deterministic data flows, while LLMs operate non-deterministically, especially when using AI agents and tool calls~\cite{liu2024formalizing}, and furthermore from the difficulties that many governing bodies face with trying to regulate this fast-moving and dynamic new research paradigm. Faced with a lack of clear guidance and regulation, many researchers opt to use AI tools heavily -- and sometimes irresponsibly -- in order to gain a competitive advantage.

\subsection{Privacy-utility Trade-off}
% : Conflict Between Institutional Compliance and Academic Priority

The intense ``publish-or-perish'' culture creates a severe conflict between strict institutional regulations, such as IRB protocols, and the competitive advantages of AI tools. To secure data, university policies typically mandate localized models or rely on user restraint. However, these approaches fail by ignoring researchers' extreme productivity pressures and the limitations of self-hosted architectures. 
Consequently, expecting user restraint is an unsustainable security posture. The high functional payoff of AI forces researchers to accept privacy risks to maintain research output~\cite{tolsdorf2025safety}. Abstaining from AI creates such a severe competitive disadvantage that users routinely rationalize data exposure as an unavoidable operational cost. This prioritization of efficiency over compliance is universal: even experienced privacy analysts processing hyper-sensitive data prefer LLM assistance to reduce effort~\cite{kramer2025integrating}. Therefore, institutional mandates that simply prohibit cloud-based AI may be bypassed in everyday research practices.

\subsection{Implications}

To address the aforementioned vulnerabilities, we propose implications across three levels: data control, interface transparency, and socio-technical aspects.

\textbf{Data Control} \textit{Separated conversational workspaces.} To balance data privacy with system usefulness, architectures could separate memory-based and unlogged interactions. Institutions could provide tools similar to Harvard's ``AI sandbox''~\cite{harvard_ai_sandbox_gazette} or NUS's ``AI-Know''~\cite{nus_ai_know_history}, which offers educational environments that keep data entirely within the institution. This could be paired with customized interfaces, with one secure side for sensitive data, and another side with low-stakes tasks like literature reviews. This separation protects important research data from being stored by external service providers.

\textit{Automated data screening.} IRBs could use automated screening tools beyond mere checklists, similar to the privacy-compliance monitoring tools in corporations~\cite{rutgers_interactive_irb_tools}. These tools monitor and flag sensitive research data in user prompts or uploaded files before they are sent to chatbots. However, designers must carefully balance these protections to respect academic freedom and avoid creating invasive academic surveillance~\cite{kwapisz2024privacy,stegman2023my}.

\textbf{Interface Transparency} \textit{Data usage transparency.} Systems should provide real-time visual signs showing exactly how specific inputs are handled, which clearly separate data kept for model training, temporarily saved data, and completely unlogged chats. This transparency reduces the efforts needed to manually hide sensitive data and helps researchers make informed privacy-preserving choices.

\textbf{Socio-technical Aspects} \textit{Interactive privacy training.} Static rule checklists are often difficult to apply on LLMs, whose data flow is opaque and dynamic~\cite{harvard_hms_genai_guidance}. Institutions should clearly distinguish between educational usage and research usage. Research-oriented training should adopt interactive, scenario-based tutorials. By visually showing how adversarial prompts or other attacks can accidentally leak a researcher's core ideas, systems can explain how institution-provided tools mitigate these risks~\cite{harvard_catalyst_ai_educational_resources,oxford_genai_research_faq}.

\textit{Dedicated channel for AI risks.} Novice researchers should have a dedicated channel for AI-related research risks, which unite different perspectives such as IRB, IT and IP offices and provide comprehensive, actionable plans. This entry points should have clear guidance for varying data types, such as diverse information levels, and mappings to AI usage restrictions, like those in the Harvard's models~\cite{harvard_hms_genai_guidance,harvard_catalyst_etech}. 

\textit{Promoting academic privacy champions.} Learning from successful corporate examples~\cite{tahaei2021privacy}, universities should identify and support privacy champions among novice researchers. Tools could include community reporting or template sharing to encourage peer-led privacy-preserving habits. However, these programs must be applied carefully to avoid adding administrative burdens that reduce researchers' motivation~\cite{rutgers_irb_toolkit}.

\section{Limitations and Future Work}

We acknowledge several limitations in this paper. First, the reliance on convenience sampling via online platforms introduces selection bias. While we recruited a diverse pool spanning multiple countries and backgrounds, the findings, especially code frequencies, may not fully generalize to the global research population. Future research should use large-scale longitudinal studies to observe the evolution of user mental models. Second, as with most qualitative studies, our data is subject to self-reporting biases, such as recall bias and social desirability bias. To mitigate this, future work should include empirical technical audits to objectively measure the discrepancy between user-perceived mitigations and actual usage. Third, our paper focuses on novice researchers, whose workflows and risk perceptions may differ from established scholars. Comparative studies involving senior researchers could reveal how career stages and institutional responsibilities influence research risk assessments.

\section{Conclusion}

This paper examines novice researchers' privacy perceptions regarding LLM-assisted research workflows through semi-structured interviews (N=44). Findings reveal a prioritization of productivity over ideas' protection. Users rely on ad-hoc mitigation strategies, such as data fragmentation, but feel that they serve merely as psychological placebos against backend retention. A critical discrepancy exists between users' mental models, which falsely assume input dilution prevents memorization, and actual technical risks. We therefore recommend actions such as institutional automated screening and separated tools, transparent risk visualization, and interactive privacy training..

% This paper provides a qualitative investigation (N=44) into the research privacy landscape of novice researchers in the era of generative AI. Our findings indicate that researchers perceive these tools as an operational necessity and prioritize productivity over strict data protection even when they recognize risks such as scooping and intellectual property theft. Although participants adopt diverse mitigation strategies including data sanitation and fragmentation, they often perceive these actions as psychological placebos that provide limited security against backend data retention. We identified a specific discrepancy between user mental models and technical risks, where researchers rely on the assumption that input dilution and manual obfuscation prevent data memorization. The results highlight a persistent accountability deficit and a demand for transparent data governance that provides granular control over the data lifecycle. These insights suggest that institutional guidelines must address the non-deterministic nature of generative AI to better protect the integrity of scholarly work.

\begin{acks}
We acknowledge the use of Gemini 3.1 Pro and ChatGPT strictly for minor editing, specifically grammar and style polishing. Authors retain full responsibility for the accuracy, originality, and integrity of this paper.
\end{acks}

\bibliographystyle{ACM-Reference-Format}
\bibliography{main}

\appendix
\clearpage

\section{Ethical Considerations}

We acknowledged that our paper has ethical concerns. We adhered to the Menlo report~\cite{bailey2012menlo} and the Belmont report~\cite{beauchamp2008belmont} in mitigating the ethical concerns, and all interviews in our paper acquired the approval of our university's Institutional Review Board (IRB).

Following the principle of Respect for Persons, we obtained informed consent from all researchers participating in our interviews. Participants were briefed on the study's objectives, and retained the right to withdraw or delete their data at any time without any reasons. We prioritized the mitigation of risks associated with the disclosure of sensitive research workflows and potential institutional non-compliance. Given that novice researchers often operate within highly competitive research environments, revealing their reliance on commercial LLMs carries inherent professional risks. To protect participants from potential institutional retribution or academic scooping, we rigorously anonymized all PIIs, detailed research topics, and proprietary methodologies. Furthermore, we highlighted that the observed practices and perceptions originate from multiple constraints rather than individual negligence, underscoring the need for multi-stakeholder interventions to mitigate these risks.
% We highlighted that the participants' practices and mental models resulted from a combination of constraints, rather than the issues of researchers themselves, which required efforts from different stakeholders to mitigate the risks.

To uphold the principles of Beneficence and minimize risks associated with the exposure of sensitive data, we strictly anonymized all PII and sensitive research workflows shared by the participants.

Furthermore, because our interview explicitly investigates the acceptability of inputting research ideas into LLMs, we applied the Menlo Report's guidelines, regarding responsible data management in information communication technology research. Consequently, to prevent any potential reverse-engineering or exploitation of our participants' specific research privacy vulnerabilities, we deliberately chose not to open-source the original interview transcripts. This decision ensures that we protect the participants' professional safety and intellectual property while still sharing our interview questions and codebooks to facilitate reproducibility.

% \section{Open Science}

% We facilitated reproducibility through open-sourcing our interview scripts, questionnaires and codebooks. However, due to ethical concerns, we choose not to open-source the original interview transcripts and the questionnaire results. The materials are provided in \href{https://anonymous.4open.science/r/CCS26B-28BE/}{https://anonymous.4open.science/r/CCS26B-28BE/} and would be published upon acceptance.

% \section{Usage of Generative AI}

% % We declared our usage of generative AI in this paper. We solely used Gemini-3.1-pro and ChatGPT (i.e., GPT-5.2) for polishing the paper. All LLM-edited text are scrutinized by authors. All the intellectual content are from the authors. The authors hold the responsibility towards all text.
% We declare the use of LLMs, specifically Gemini-3.1-pro and ChatGPT (GPT-5.2), exclusively for linguistic refinement and manuscript polishing. All original intellectual content, including research design, data analysis, and conceptual frameworks, was produced solely by the authors. The authors have meticulously reviewed and verified all AI-assisted revisions and assume full responsibility for the integrity and accuracy of the final text.

\section{Interview Scripts}
\label{tab:interview}

The following is the interview scripts. We used Chinese or English to interview participants according to participants' preference. A primary author who is fluent in both English and Chinese translated the script to Chinese, and the other authors checked the interview script to ensure its correctness.

\subsection{Basic Usage Experience}

1. Which generative AI tools or LLMs do you primarily use in your research flow?

2. Could you describe your primary use cases for these tools? (prompt if needed, e.g., data analysis, drafting, coding)

3. How do you typically input your data into these systems? (prompt id needed, e.g., direct text entry, file uploads) What drives this preference?
% Do you usually copy-paste text, or do you upload entire PDFs/code files to the bot? Why?

\subsection{Privacy Perceptions \& Concerns}

4. What do you believe happens to your data after you submit a prompt. Where does the input go, and how is it processed?

5. What is your understanding of the data retention policies of the tools you used? (explain if needed, e.g., do you believe the data is permanently stored, used for secondary purposes, or deleted)

6. Are you aware of any built-in privacy controls offered by these platforms? (If yes: how, if at all, do you use them?)

7. How do you perceive the privacy of the research data you share with these platforms?

8. Are there specific categories of research data or tasks you withhold from AI platforms? If so, what are they and why?

9. Where do your privacy perceptions originate? 

\subsection{Specific Experience Exploration}

10. Could you share a specific instance where you felt the privacy risk was highest, or that left the deepest impressions on you?

11. (If applicable) What exactly happened? (What did you input? What was the AI's response?)

12. (If applicable) Why did it feel ``risky'' at that moment?

13. (If applicable) How did you handle it afterward? (Deleted the chat? Took remedial actions?)

14. (If applicable) Did this experience change how you use AI later on?

15. How do you perceive the risks that your inputs might be used to train the model, potentially leaking your research ideas to other users?

16. When using AI for peer review, how do you perceive that AIs might identify the author or leak your identity as a reviewer?

17. Does the geographical origin (e.g., domestic vs. international) affect your trust in these tools? Why?

\subsection{Coping Strategies}

18. What strategies,if any, do you employ to protect research data before or during interaction with LLM-assisted tools?

19. How would you perceive the privacy communication of LLM-assisted tools (e.g., privacy policy)?

% 17. Do you sanitize your input? (e.g., replacing specific formulas, removing names, or only inputting non-sensitive paragraphs)

% 18. Do you ever switch to locally hosted models or avoid AI entirely for sensitive tasks?

% 19. If a tool is highly effective but has a vague privacy policy, would you still use it? How do you balance efficiency and privacy?

% \subsection{System Understanding}

% 20. Where do you think your text goes after you click ``send''? Where is it eventually stored?

% 21. Are you aware of features like ``opt-out of training'' or ``temporary chat''? Have you used them?

\subsection{Suggestions \& Expectations}

20. If you could design a chatbot specifically for researchers, what privacy features would it have?

21. Who do you believe should bear the primary responsibility for safeguarding research data?
% Who should be responsible for research privacy: the individual, the university (e.g., via enterprise licenses), or the government (via legislation)?

22. Is there anything else regarding LLM-assisted research and privacy that you want to share?

\section{Codebook}\label{app:codebook}

Drawing upon the thematic analysis, we developed a codebook to systematically analyze novice researchers' concerns regarding LLM-based research assistants. Table~\ref{tab:codebook} presents the finalized codebook, structured by themes, corresponding codes, and operational descriptions.

\begin{table*}[htbp]
\centering
\caption{Codebook of researchers' privacy concerns, mitigation and challenges.}
\label{tab:codebook}
\begin{tabularx}{\textwidth}{>{\raggedright\arraybackslash}p{3.5cm} >{\raggedright\arraybackslash}p{4.5cm} X}
\toprule
\textbf{Theme} & \textbf{Code} & \textbf{Description} \\
\midrule
\multirow{4}{3.5cm}{\textbf{Understanding of LLM Data Handling}} 
& Opacity \& complexity & Perception of LLM data pipelines as intransparent, hindering researchers' ability to understand backend storage mechanisms. \\
& Unavoidable knowledge leakage & The fatalistic assumption that interacting with cloud-based models inevitably results in the exposure of submitted proprietary data. \\
& Iterative optimization & The belief that user prompts and proprietary data are used to continuously train and refine the underlying models. \\
& Obfuscation by AIs & The perception that AI platforms deliberately mask their data harvesting practices through convoluted interfaces and vague policy language. \\
\midrule
\multirow{8}{3.5cm}{\textbf{Perceived Privacy Risks}} 
& Information categories that matter & Differentiation of risk severity based on data types, with strict concerns on unpublished manuscripts and raw datasets. \\
& Unauthorized disclosure of intellectual property & Anxiety over the exposure of novel hypotheses and core research ideas to competing users or the public. \\
& Empirical data exposure & Fears concerning the leakage of primary datasets, including sensitive human-subject information or proprietary experimental metrics. \\
& Deanonymization and profiling & Risks of models aggregating query histories to infer researchers' identities, institutional affiliations, or distinct behavioral profiles. \\
& Memorization and data deletion & Apprehensions that LLMs permanently retain inputs in their weights, rendering user-initiated data deletion requests ineffective. \\
& Security vulnerabilities & Concerns regarding centralized data breaches or unauthorized access by third-party actors exploiting AI platform weaknesses. \\
& Negligible risks & Users perceive the utility of the LLM to outweigh the potential privacy implications. \\
& Research vs. daily privacy risks & The contextual distinction between the rigorous safeguarding of high-stakes academic assets versus relaxed attitudes toward routine, non-sensitive interactions. \\
\midrule
\multirow{4}{3.5cm}{\textbf{Knowledge Sources}} 
& Formal pedagogical and institutional instruction & Privacy mental models shaped by university guidelines, ethical compliance training, and official organizational policies. \\
& Empirical engagement and experiential discovery & Understanding derived from direct, hands-on interaction with LLMs and longitudinal observation of their output behaviors. \\
& Peer-mediated and collaborative discourse & Informal discussions, shared workflows, and warnings among academic colleagues. \\
& Scholars and medias & Published security research, expert commentary, and public media reports. \\
\midrule
\multirow{6}{3.5cm}{\textbf{Privacy Protection Practices}} 
& Architectural isolation and access control & Using localized, offline models or enterprise-tier accounts to ensure data and access boundaries. \\
& Data sanitation and obfuscation & Manually redacting identifiable metrics, substituting sensitive terminology, or applying masking techniques prior to input. \\
& Fragmentation and decoupled processing & Deconstructing complex queries into disjointed, context-free segments to prevent the model from comprehending the holistic idea. \\
& Adversarial testing, probing and deletion & Actively testing the model's memorization limits and routinely purging chat histories to minimize persistent exposure windows. \\
& Human-centric verification and awareness & Maintaining vigilant oversight of AI outputs and cross-referencing generated content to detect potential privacy anomalies. \\
& Ineffectiveness of current mitigation & The recognition that existing user-side privacy strategies are often insufficient. \\
\midrule
\multirow{4}{3.5cm}{\textbf{Challenges}} 
& Privacy-publication trade-off & Tension between utilizing AI to accelerate publication and risk of exposing novel findings. \\
& Intransparency and informational asymmetry & The imbalance of knowledge where AI providers conceal data practices and retention policies from user. \\
& Lack of control & Absence of mechanisms for users to actively manage, restrict, or revoke access to their submitted data. \\
& Accountability deficit & Difficulty in assigning liability or tracing when research is inadvertently leaked. \\
\bottomrule
\end{tabularx}
\end{table*}

\onecolumn
\section{Participants' Demographics}

Table~\ref{tbl:participant} showed the participants demographics.

\begin{table*}[h]
\centering
\caption{Participants' demographics. For position, if the participants are currently pursuing PhD or master, we denoted PhD or master, or if they already became faculties or are currently research assistants, we denoted correspondingly.}
\small
\label{tbl:participant}
\begin{tabularx}{1\textwidth}{ccp{1cm}p{1.5cm}p{2.2cm}p{2cm}p{3.3cm}X}
\hline
\textbf{ID} & \textbf{Age} & \textbf{Gender} & \textbf{Channel} & \textbf{Position} & \textbf{Nationality} & \textbf{Country of Current Study} & \textbf{Research direction} \\ \hline
1 & 27 & F & WeChat & PhD & Mainland China & Mainland China & Architecture \\ \hline
2 & 22 & F & WeChat & PhD & Mainland China & Mainland China & Life Sciences \\ \hline
3 & 23 & F & Contact & Master & Mainland China & Mainland China & Human-Computer Interaction \\ \hline
4 & 31 & F & WeChat & PhD & Mainland China & Mainland China & Business Administration \\ \hline
5 & 28 & F & WeChat & PhD & Mainland China & Mainland China & Biological Science, Bioengineering \\ \hline
6 & 30 & M & RedBook & PhD & Mainland China & Mainland China & Information Management \\ \hline
7 & 25 & F & RedBook & PhD & Mainland China & Netherlands & Sociology \\ \hline
8 & 28 & M & RedBook & PhD & Mainland China & UK & Computational Social Science \\ \hline
9 & 24 & F & RedBook & PhD & Mainland China & USA & Human-Computer Interaction \\ \hline
10 & 25 & F & RedBook & Master & Mainland China & USA & Materials Science \\ \hline
11 & 23 & F & RedBook & Master & Mainland China & Hong Kong & Literature \\ \hline
12 & 26 & F & RedBook & PhD & Mainland China & Canada & Finance and Economics \\ \hline
13 & 27 & F & RedBook & PhD & Mainland China & Netherlands & Psychology \\ \hline
14 & 25 & M & RedBook & Master & Mainland China & UK & Artificial Intelligence \\ \hline
15 & 26 & F & RedBook & PhD & Mainland China & USA & Educational Psychology \\ \hline
16 & 30 & F & RedBook & Professor & Mainland China & Mainland China & Toxicology \\ \hline
17 & 22 & M & RedBook & Master & Mainland China & USA & Statistics \\ \hline
18 & 25 & M & RedBook & PhD & Mainland China & Germany & Linguistics \\ \hline
19 & 26 & F & RedBook & PhD & Mainland China & USA & Psycholinguistics \\ \hline
20 & 25 & F & RedBook & PhD & Mainland China & USA & Economics and Management \\ \hline
21 & 25 & M & RedBook & PhD & Mainland China & USA & Mechanical Engineering \\ \hline
22 & 28 & M & RedBook & Postdoc/Faculty & Mainland China & Mainland China & Social Sciences \\ \hline
23 & 29 & F & RedBook & PhD & Mainland China & Mainland China & Humanities and Social Sciences \\ \hline
24 & 28 & F & LinkedIn & PhD & Sri Lanka & Australia & HCI \\ \hline
25 & 28 & M & RedBook & PhD & Mainland China & Germany & Law \\ \hline
26 & 28 & M & RedBook & PhD & Mainland China & Mainland China & Public Administration \\ \hline
27 & 26 & M & RedBook & PhD & Mainland China & Mainland China & Education \\ \hline
28 & 27 & F & RedBook & PhD & Mainland China & UK & Education \\ \hline
29 & 28 & M & RedBook & Corporate Researcher & Mainland China & USA & Biological Sequencing \\ \hline
30 & 25 & F & RedBook & PhD & Mainland China & USA & Media Arts \\ \hline
31 & 26 & F & Campus & PhD & Malaysia & Mainland China & Medical Image Processing \\ \hline
32 & 37 & F & Campus & PhD & India & Mainland China & Chinese Linguistics \\ \hline
33 & 38 & M & LinkedIn & Lecturer & Indonesia & Indonesia & Human-Computer Interaction \\ \hline
34 & 24 & F & Campus & PhD & Ghana & Mainland China & Chinese Literature \\ \hline
35 & 28 & F & Campus & Master & Bangladesh & Mainland China & Teaching Chinese to Speakers of Other Languages \\ \hline
36 & 27 & M & Campus & PhD & Mainland China & Finland & Auditing \\ \hline
37 & 24 & M & Campus & Master & Pakistan & Mainland China & Chinese Education \\ \hline
38 & 23 & M & WeChat & PhD & Singapore & Singapore & Chinese Language \\ \hline
39 & 31 & F & Campus & Master & Malaysia & Mainland China & Linguistics \\ \hline
40 & 28 & F & Campus & Master & Sri Lanka & Mainland China & Economics and Management \\ \hline
41 & 27 & F & Campus & Master/RA & Pakistan & Mainland China & Artificial Intelligence \\ \hline
42 & 28 & M & Campus & PhD & Pakistan & Mainland China & AI-driven Origami Structures \\ \hline
43 & 23 & M & Campus & Master & Uzbekistan & Mainland China & Chinese Language \\ \hline
44 & 27 & F & Campus & Master & Indonesia & Mainland China & Teaching Chinese as a Second Language \\ \hline
\end{tabularx}
\end{table*}

\end{document}